# Strong magnetic instability in correlated metal $Bi_2Ir_2O_7$

T. F. Qi[1], O. B. Korneta[1], and Xiangang Wan[2], L. E. DeLong[1], P. Schlottmann[3], and G. Cao[1*]

[1]Department of Physics and Astronomy and Center for Advanced Materials, University of Kentucky, Lexington, KY 40506, USA

[2] National Laboratory of Solid State Microstructures and Department of Physics, Nanjing University, Nanjing, China

[3]Department of Physics, Florida State University, Tallahassee, FL32306, USA

The interplay of spin-orbit interactions and electronic correlations dominates the physical properties of pyrochlore iridates, $R_2Ir_2O_7$ (R = Y, rare earth element), which are typically magnetic insulators.  We report an experimental/theoretical study of single-crystal $Bi_2Ir_2O_7$ where substitutions of Bi for R sensitively tips the balance between competing interactions so as to favor a metallic state with a strongly exchange enhanced paramagnetism. The ground state is characterized by the following features: **(1)** A divergent low-temperature magnetic susceptibility that indicates no long-range order down to 50 mK; **(2)** strongly field-dependent coefficients of the low-temperature T- and $T^3$-terms of the specific heat; **(3)** a conspicuously large Wilson ratio $R_W \approx 53.5$; and **(4)** unusual temperature and field dependences of the Hall resistivity that abruptly change below 80 K, without any clear correlation with the magnetic behavior.  All these unconventional properties suggest the existence of an exotic ground state in $Bi_2Ir_2O_7$.

PACS: 75.40.-s, 75.70.Tj, 71.20.-b

I. Introduction

Iridates have recently proved to be a fertile ground for studies of new physics driven by the competition between spin-orbit coupling (SOC) and the electron-electron Coulomb interaction U. The new physics is manifest in a large array of novel experimental phenomena, such as the $J_{eff} = 1/2$ Mott state [1-3], lattice-driven magnetoresistivity [4], giant magnetoelectric effects [5], unusual orbital magnetism [6], a spin liquid phase in a hyper-kagome structure [7], and a zig-zag magnetic structure [8, 9]. These highly unusual observations have been accompanied by a growing list of related theoretical proposals and predictions: high-$T_c$ superconductivity [10], Weyl semimetals with Fermi arcs [11], Axion insulators with strong magnetoelectric coupling [12], topological insulators [13], correlated topological insulators with large gaps enhanced by Mott physics [14, 15], Kitaev modes [16], 3-D spin liquids with Fermionic spinons [17], and topological semimetals [18]. Indeed, the relativistic SOC, which is proportional to $Z^4$ (Z = atomic number), cannot be treated as a perturbation in the case of iridates. Instead, the SOC (magnitude 0.4 − 1 eV) strongly competes with U (magnitude 0.4 − 2.5 eV) and certain other interactions, which critically biases their mutual competition in order to stabilize ground states with exotic physical properties.

Traditional arguments suggest that the iridates should be more metallic and less magnetic than materials based upon 3*d* and 4*f* elements because 5d-electron orbitals are more extended in space, which increases electronic bandwidth. This conventional wisdom sharply contrasts with two conspicuous characteristics commonly observed among layered iridates: **(1)** relatively high-temperature, weak ferromagnetism; and **(2)** insulating states with exotic behavior [4-6, 19-25]. It is now recognized that strong SOC can drive novel narrow-gap Mott insulating states in iridates [1,2]. For example, strong crystal fields split 5d band states with $e_g$ symmetry in $Sr_2IrO_4$, and $J_{eff} = 1/2$ and $J_{eff} = 3/2$ multiplets arises from $t_{2g}$ bands via SOC. A weak admixture of the $e_g$ orbitals downshifts the $J_{eff} = 3/2$ quadruplet from the $J_{eff} = 1/2$ doublet. Since the $Ir^{4+}$ ($5d^5$) ions provide five electrons, four of them fill the lower $J_{eff} = 3/2$ bands, and one electron partially fills the $J_{eff} = 1/2$ band. The $J_{eff} = 1/2$ band is so narrow that even a reduced U is sufficient to open a small gap supporting a traditionally unexpected insulating state.

The pyrochlore iridates, $R_2Ir_2O_7$ (R = Y or rare earth element) are magnetic insulators, except for $Pr_2Ir_2O_{7-\delta}$, which exhibits correlated metallic behavior [26-30]. These iridates have been intensively studied, since they display a variety of novel physics. It is established that the ground state of these materials sensitively depends on the relative strengths of competing SOC and U, and a hybridization interaction that is controlled by the Ir-O-Ir bond angle [11-13]. Therefore, small perturbations, such as changes in lattice parameters caused by variations of the ionic radius of different R ions, can easily tip the balance between the competing energies and ground states [30].

However, the generally large and stable magnetic moments carried by the rare earth elements can mask the intriguing physics generated by the 5d electrons of the Ir ions in $R_2Ir_2O_7$, in which the $Ir^{4+}$ and $R^{3+}$ ions form two interpenetrating FCC sublattices [31]. We have therefore mounted a combined experimental and theoretical investigation of single-crystal $Bi_2Ir_2O_7$ where substitution of the $Bi^{3+}$ ion for the rare earth significantly enhances the hybridization between the Bi-6s/6p and Ir-5d electrons, which overpowers the SOC and U, driving the material into a metallic state. This metallic state is characterized by the following unconventional features: **(1)** The magnetic susceptibility diverges at low temperatures, and the specific heat C(T) exhibits no anomalies that would indicate long-range order down to 50 mK. **(2)** The coefficients of the T- and $T^3$-terms of C(T) are strongly field-dependent and are enhanced with increasing magnetic field H at low temperatures. **(3)** The Wilson ratio $R_W$ reaches 53.5 at T = 1.7 K and $\mu_oH = 0.5$ T, far beyond typical values for highly correlated systems (e.g., $R_W$ ~1-6), but comparable to those observed for $SrIrO_3$ **[32]**. **(4)** The Hall resistivity $\rho_H(T)$ drops abruptly below 80 K in modest applied fields H, which has no clear relevance to the magnetic behavior. Our electronic structure calculations indicate a metallic ground state with a Fermi energy $E_F$ near a sharp peak in the density of states (DOS) *g($E_F$)*, qualitatively captures the curious physical properties observed in $Bi_2Ir_2O_7$.

## II. Experimental Techniques

Measurements of magnetization M(T,H), heat capacity C(T), electrical resistivity $\rho(T)$ and Hall resistivity $\rho_H(T,H)$ were performed over an extended temperature interval, 50 mK < T < 400 K, using either a Quantum Design (QD) Physical Property Measurement System or QD Magnetic Property Measurement System equipped with a Linear Research Model 700 AC bridge. Single crystals were grown from off-stoichiometric quantities of $IrO_2$ and $Bi_2O_3$ using self-flux techniques in an oxygen rich atmosphere. Similar technical details are described elsewhere **[18-23]**. The crystal structure was determined from a small fragment of a single crystal using Mo *Kα* radiation and a Nonius Kappa CCD single-crystal diffractometer at 90 K and 295 K. The structures were refined using the SHELX-97 programs **[33, 34]**. The composition of the crystals was examined by energy-dispersive X-ray (EDX) spectroscopy. Crystal quality is evaluated by looking at single crystal x-ray diffraction data quality parameters such as mosaicity, cell parameter standard deviations, resolution, **R**-factors, and others. The mosaicity of the single crystals studied are small, which indicates well-ordered crystals and hence better diffraction. The standard deviations of all lattice parameters and interatomic distances are smaller than 0.1%.

### III. Crystal Structure and Stoichiometry

The crystal structure of $Bi_2Ir_2O_7$ adopts a cubic cell with space group *Fd3m* (No. 227) and lattice parameter $a$ = 10.3111(12) Å; the Bi ion occupies the 16*d* site (1/2,1/2,1/2), the Ir ion the 16*c* site (0,0,0), O1 the 48*f* site (0.3325, 1/8,1/8), and O2 the 8*b* site (3/8,3/8,3/8). This structure is similar to that reported previously **[35].** Our single-crystal x-ray diffraction at 90 K and 295 K also yields a converged refinement for a structure having Bi at the 16*c* site and Ir at the 16*d* site. But a close examination of the refinement data (see details in **Table 1**) reveals that refinement parameters that measure the quality of refinements, $R[F^2>2\sigma(F^2)]$ factor, $wR(F^2)$ (weighted R factor), and goodness of fit *S*, uniformly favor the former structure with the Ir ion at the 16*c* site. (The smaller these parameters are, the better refinements are.) In addition, the maximum and minimum difference map features $\Delta\rho_{max}$, $\Delta\rho_{min}$ in **Table 1** are also small, which is further evidence of high-quality crystal structure. These parameters constitute the direct evidence for the validity of the structure refinement. It is also important to examine the oxygen content in the pyrochlores as it could vary. We therefore checked *atom site occupancy factor* on each oxygen site by running a test in which the oxygen occupancies were allowed to refine freely. They all refined to values greater than 1, which suggests there is no discernible oxygen vacancy found from our refinement, confirming the refined stoichiometry being $Bi_2Ir_2O_7$.

**Table 1** *Experimental and refinement details of two different refinements at 90K*

|  | Ir at *16c* | Bi at *16d* |
| --- | --- | --- |
| Crystal data | | |
| **Chemical formula** | $Bi_2Ir_2O_7$ | $Bi_2Ir_2O_7$ |
| **$M_r$** | 1828.72 | 1828.72 |
| **Crystal system, space group** | Cubic, *Fd-3m* | Cubic, *Fd-3m* |
| **Temperature (K)** | 90K | 90K |
| ***a* (Å)** | 10.3111(12) | 10.3111 |
| **α, β, γ (°)** | 90, 90, 90 | 90, 90, 90 |
| **V (Å$^3$)** | 1096.3 (2) | 1096.3 (2) |
| **Z** | 4 | 4 |
| **Radiation type** | Mo *Kα* | Mo *Kα* |
| **μ (mm$^{-1}$)** | 112.40 | 112.40 |
| **Crystal size (mm)** | 0.05 ×0.05 ×0.05 | 0.05× 0.05 × 0.05 |
| Data collection | | |
| **Diffractometer** | Nonius KappaCCD diffractometer | Nonius KappaCCD diffractometer |
| **Absorption correction** | Multi-scan SADABS(Sheldrick, 1996) | Multi-scan SADABS(Sheldrick, 1996) |
| **No. of measured, independent and observed[$I> 2\sigma(I)$] reflections** | 5638, 83, 68 | 5638, 83, 68 |
| **$R_{int}$** | 0.052 | 0.052 |
| Refinement | | |
| **$R[F^2>2\sigma(F^2)]$, $wR(F^2)$, $S$** | 0.019, 0.046, 1.14 | 0.021, 0.053, 1.23 |
| **No. of reflections** | 83 | 83 |
| **No. of parameters** | 10 | 10 |
| **No. of restraints** | 0 | 0 |
| **$\Delta\rho_{max}$, $\Delta\rho_{min}$ (e Å$^{-3}$)** | 1.15, -1.35 | 1.34, −1.87 |

## IV. Results and Discussions

One of the distinct characteristics of $Bi_2Ir_2O_7$ is that its DC magnetic susceptibility χ(T) exhibits no evidence of long-range order, but appears to diverge below 30 K, indicating a proximity to a magnetic instability, as shown in **Fig. 1a**. Fitting χ(T) for T < 50 K to a Curie-Weiss law defined by $\chi = \chi_o + c/(T-\theta_{CW})$ (see

**Inset** in **Fig.1a**) gives rise to the temperature-independent susceptibility $\chi_o$, the Curie constant c and the Curie-Weis temperature $\theta_{CW}$, which are tableted in **Table 2**. The unusually large $\chi_o$ (=1.07 x $10^{-2}$) indicates that a strongly exchange enhancement, $Ug(E_F)$, is present in $Bi_2Ir_2O_7$. In contrast, $IrO_2$ is a Pauli paramagnet with a much smaller, temperature-independent $\chi$ (~$10^{-4}$ emu/mole-Ir) **[32]**. A nearly negligible $\theta_{CW}$ is consistent with the absence of the long-range order (**Table 2**). In addition, the isothermal magnetization M(H) is readily saturated in a modest applied field $\mu_oH$ ~ 2 T, whereas the extrapolated saturation moment $\mu_S$ is ~ 0.013 $\mu_B$/Ir at T = 1.7 K, which is less than 2% of that expected (1 $\mu_B$/Ir) for a S = 1/2 system. $\mu_S$ rapidly decreases as T is raised from 1.7 K to only 5 K, as seen in **Fig. 1b**, which reaffirms the possibility of the existence of a magnetic instability at lower temperatures. Indeed, a large ratio of $\mu_{eff}/\mu_S$ (= 7.7) with $T_C$ ~ 0 places $Bi_2Ir_2O_7$ in a region characterized by strong spin fluctuations in the Rhode-Wohlfarth plot **[36]**.

**Table 2** *Estimated magnetic parameters from **Fig.1***

| $\chi_o$ (emu/mole) | $\theta_{CW}$ (K) | $\mu_S$ ($\mu_B$/Ir) | $\mu_{eff}$ ($\mu_B$/Ir) | $\mu_{eff}/\mu_S$ |
|---|---|---|---|---|
| 1.07 x $10^{-2}$ | - 2.3 | 0.013 | 0.10 | 7.7 |

Consistent with $\chi(T)$, the specific heat C(T) shows no evidence for a long-range order down to 50 mK, as shown in **Fig. 2**. Note that the sharp upturn (see **Fig. 2 inset**) is due to a strong Schottky effect that dominates C(T) for T < 100 mK. It is notable that dC(T)/dT shows a downturn below 30 K (see **Fig. 2a**), which tracks $\chi(T)$ in the same temperature range (**Fig. 1a**), indicating a magnetic crossover occurring near 30 K.

A close examination of the lower temperature C(T,H) reveals intriguing insights to the ground state of $Bi_2Ir_2O_7$. As shown in **Fig. 2b**, C(T) for 50 mK < T < 4 K is well described by $C(T) = \gamma T + \beta T^3$, where $\gamma$ measures the effective mass m* or the DOS near $E_F$, and β measures the phonon and/or magnetic contributions. The data in **Figs. 2b** and **2c** clearly illustrate that both γ and β are sensitive to applied magnetic field H. Indeed, γ rises from 16 mJ/mole $K^2$ at $\mu_o H = 0$ to 25 mJ/mole $K^2$ at $\mu_o H = 14$ T. This field-induced increase in γ indicates that renormalizations of m* are significant and of magnetic origin, and that the normalized bandwidth is so narrow that it is sensitive to even modest magnetic fields. Such behavior is consistent with our band calculations in which the Fermi energy resides near a sharp peak of the DOS discussed below.

Conventionally, the field-independent phonon contribution dominates the term $\beta T^3$; this is also true for $Bi_2Ir_2O_7$ at high temperatures, and a fit of the data to $C(T) = \gamma T + \beta T^3$ for 20 < T < 100 K yields a Debye temperature, $\theta_D = 232$ K, which is smaller but comparable to those of other iridates such as $SrIrO_3$ **[32]**. It is therefore striking that the low-T coefficient β extrapolated in 50 mK < T < 4 K increases with H ten times more rapidly than γ does, rising from 0.97 mJ/mole $K^4$ at $\mu_o H = 0$ T, to 12.9 mJ/mole $K^4$ at $\mu_o H = 14$ T. The fact that both γ and β simultaneously increase with H suggests that the β is also dominated by electronic rather than phonon degrees of freedom (**Fig. 2c**). The spin-fluctuations, which should be antiferromagnetic short-range correlations, have a linear dispersion and a broad linewidth (also increasing with k) at small k; and it is the linear dispersion that gives rise to the $T^3$-term. The negative and small $\theta_{CW}$ (= - 2.3 K) along with the weak $\mu_S$ (**Table 2**) suggest a very weak antiferromagnetic coupling; therefore a modest magnetic field is strong enough to quench the antiferromagnetic correlations and push entropy to lower temperatures. This explains the increases in **(a)** χ(T), **(b)** γ and **(c)** β. A field-dependent β is rare as a conventional magnetic state is robust enough to energetically stable in the presence of modest magnetic fields (say, 14 T, which amounts to no more than 1.5 meV). But the field-dependence of β is not entirely without precedent; a similar yet weaker field dependence of β is recently observed in a correlated cobalt oxide due to a magnetic instability [37].

Indeed, the Wilson ratio, $R_W \equiv \pi^2 k_B^2 \chi / 3\mu_B^2 \gamma$, which measures the relative enhancements of the spin susceptibility and electronic specific heat, reaches 53.5 at T = 1.7 K and $\mu_o H$ = 0.5 T, as shown in **Fig. 2d**. This strikingly large $R_W$, which is clearly a consequence of non-Fermi liquid behavior, is far beyond the values (e.g., $R_W$ ~1-6) typical of heavy Fermi liquids [38]. A similarly large $R_W$ value is also seen in SrIrO$_3$, where it is ascribed to non-Fermi liquid behavior in proximity to a quantum critical point [32]. However, $R_W$ drops rapidly with H to a value of 3.4 at $\mu_o H$ = 7 T, which may be related to the highly unusual increase of the β coefficient with field.

The presence of the non-Fermi liquid is further corroborated by the temperature dependence of the **a**-axis resistivity ρ (shown in **Fig. 3a**), which exhibits a $T^{3/2}$ law for 1.7 ≤ T ≤ 10 K (see **Fig. 3a inset**) with a residual resistivity $\rho_o$ = 0.34 mΩ cm and a residual resistance ratio RRR ≈ 3. The strong negative curvature of ρ(T) is typical of strongly correlated electrons, and the $T^{3/2}$ law is typical of spin fluctuation systems where it is associated with diffusive electron motion caused by strong interactions between itinerant electrons and critically-damped, very-long-wavelength magnons [39]. Electron scattering is highly sensitive to temperature, and accordingly, the power law of ρ varies widely with increasing T, changing from $T^{3/2}$ to $T^2$ (10 K < T < 17 K), $T^5$ (17 < T < 30 K), $T^{1/2}$ (40 K < T < 80 K) and finally, linear-T (T > 80 K). There is no significant magnetoresistance for fields up to 14 T, therefore the evolution of the temperature dependence of ρ is not likely to be magnetically-driven; it could be associated with a strong anharmonic effect due to soft phonons.

In sharp contrast to $\rho$, the Hall resistivity $\rho_H$ is highly sensitive to the application of applied magnetic field at low temperatures, as shown in **Fig. 3b**. While $\rho_H$ exhibits very weak temperature dependence at high temperatures, it drops sharply below 80 K, accompanied by a sign change and strong field dependence. This behavior distinctly differs from the conventional description, $\rho_H = R_oB + \mu_oR_eM$, where $R_o$ is the ordinary Hall coefficient, $B = \mu_o[H + (1 - N)M]$ is the internal induction, N is the demagnetization factor, and $R_e$ is the extraordinary Hall coefficient. Since there is no magnetically ordered state in $Bi_2Ir_2O_7$, the extraordinary Hall effect or skew scattering and/or side jump mechanisms cannot account for the unusual temperature dependence shown in **Fig. 3b**. Similar behavior is observed in a charge density wave (CDW) $NbSe_2$, and is attributed to a drastic increase in the mean free path induced by the CDW transition **[40]**. However, the observed Hall effect in **Fig.3b** is more likely due to two bands, one electron and one hole band; and the temperature dependence of the mobilities is responsible for the temperature dependence of $\rho_H$. This behavior of $\rho_H$ is qualitatively consistent with our electronic structure calculations (see below) that indicate that $E_F$ is likely located near a sharp peak in the DOS, which implies $\rho_H$ could sensitively depend on the exact position of $E_F$. Indeed, the field dependence of $\gamma$ illustrated in **Fig. 2** confirms the existence of a significant change in the DOS near $E_F$ with applied magnetic field, which in turn affects $\rho_H$. In addition, our recent study finds that strong SOC combined with applied field can change the Ir-O-Ir bond angle, which will change the quasiparticle band located at $E_F$ and various physical properties in $Sr_2IrO_4$ **[4]**.

Our band structure calculations offer more insights into the physics of $Bi_2Ir_2O_7$. We have applied the local density approximation (LDA) to density functional theory (DFT) using the full--potential, all--electron, linear--muffin--tin--orbital (LMTO) method **[41]**. The DOS and band structure obtained from non-magnetic LDA + SOC calculations are shown in **Fig. 4**. The states near $E_F$ (from -2.5 to 1.0 eV) are mainly contributed by the Ir-5d, O-2p and Bi-6p electrons; a combined effect of SOC and the crystal field interaction forms $J_{eff} = 3/2$ and $J_{eff} = 1/2$ bands that extend from -2.5 to -0.6 eV and from -0.5 to 1.0 eV, respectively. Although both $Bi_2Ir_2O_7$ and $Y_2Ir_2O_7$ (a magnetic insulator) have similar lattice parameters and Ir-O-Ir bond angle, the $J_{eff} = 1/2$ bandwidth is considerably wider in $Bi_2Ir_2O_7$ than in $Y_2Ir_2O_7$, as shown in **Figs. 4a** and **4b.** Most remarkably, $E_F$ of $Bi_2Ir_2O_7$ is located near a sharp peak in the DOS (see **Fig. 4c**); the K, X and W points, at which the band structure show anomalies, are the origin of the peak in the DOS at the Fermi level. The peak in the DOS should be very sensitive to a Zeeman splitting and to temperature smearing. These results qualitatively explain the experimental data presented in **Figs. 1-3**. We have also utilized the LSDA + U scheme to explore the potential for a finite U to stabilize a magnetic insulating state (as observed in $R_2Ir_2O_7$) in $Bi_2Ir_2O_7$. An accurate value of U is not yet known for the pyrochlores, but hybridization and screening are expected to make U smaller in $Bi_2Ir_2O_7$ compared to 2D systems such as layered $Sr_2IrO_4$ and $Ba_2IrO_4$, where U ranges between 0.5 and 2.4 eV **[42]**. Thus, assuming that U varies between 0.5 and 1.5 eV in $Bi_2Ir_2O_7$, we find that the calculations are always convergent to a metallic state with a very weak magnetic moment (< 0.08 $\mu_B$/Ir), even when an unphysically large U ( = 2.0 eV) is assumed. Clearly, the strong hybridization between Ir-5d and Bi-6p electrons suppresses the magnetic insulating state extant in the $R_2Ir_2O_7$ compounds, and drives $Bi_2Ir_2O_7$ across a quantum critical point to nearly magnetic, itinerant ground state. Calculations including dynamical correlations **[43]** are likely to also generate a sharp peak in the DOS near $E_F$ and provide additional insights into the field sensitivity of $\gamma$, $\beta$ and $\rho_H$.

Although an itinerant state does not commonly occur in the iridates, the complex interplay between SOC and other competing energies described above strongly suggests that it will exhibit extraordinary properties when it does occur. Taken together with the extremely strong exchange enhancement of the magnetic susceptibility and non-Fermi liquid behavior observed for metallic $SrIrO_3$ **[32]**, our data for $Bi_2Ir_2O_7$ indicate that the iridates may form an interesting class of materials in which strongly competing interactions exert a strong impetus toward non-Fermi liquid states with a magnetic instability or strong exchange-enhanced paramagnetic state; and $Bi_2Ir_2O_7$ provides a compelling example and a rare model system for studies of spin-orbital systems with such an exotic state.

This work was supported by NSF through grants DMR-0856234 and EPS-0814194. XW acknowledges support from the National Key Project for Basic Research of China (Grant no. 2011CB922101 and 2010CB923404), NSFC under Grant no. 11174124 and 10974082.

*Correspondence author; cao@uky.edu

**Captions:**

**Fig.1.** **(a)** The magnetic susceptibility $\chi$ as a function of temperature at $\mu_oH= 0.5$ T for H ∥ **a**-axis; and **(b)** Isothermal magnetization M vs. H at T=1.7, 5 and 50 K for H ∥ **a**-axis (left scale) and $\chi$ at T=1.7 K (right scale). **Inset**: $\Delta\chi$ vs. T for $1.7 < T < 50$ K.

**Fig.2. (a)** The specific heat C(T) and dC/dT (right scale) as a function of temperature; **(b)** C/T vs. $T^2$ for $\mu_oH$=0, 4, 8, and 14 T for 50 mK $\leq$ T $\leq$ 4K; **(c)** the field dependence of $\gamma$ and $\beta$ (right scale), and **(d)** the Wilson ratio $R_W$. $R_W$ is estimated based on $\chi$ and C/T at the corresponding field. **Inset**: C vs T for 50 mK $\leq$ T $\leq$ 4K.

**Fig.3.** The temperature dependence of **(a)** the **a**-axis electrical resistivity $\rho$, and **(b)** the Hall resistivity $\rho_H$ at $\mu_oH = 3$ and 5 T. **Inset**: the a-axis $\rho$ vs $T^{3/2}$ for $1.7 < T \leq 10$K.

**Fig.4.** Electronic structure and density of states (DOS) from LDA+SOC calculations: **(a)** Band structure of $Bi_2Ir_2O_7$; **(b)** Band structure of $Y_2Ir_2O_7$ for comparison; and **(c)** DOS of $Bi_2Ir_2O_7$.

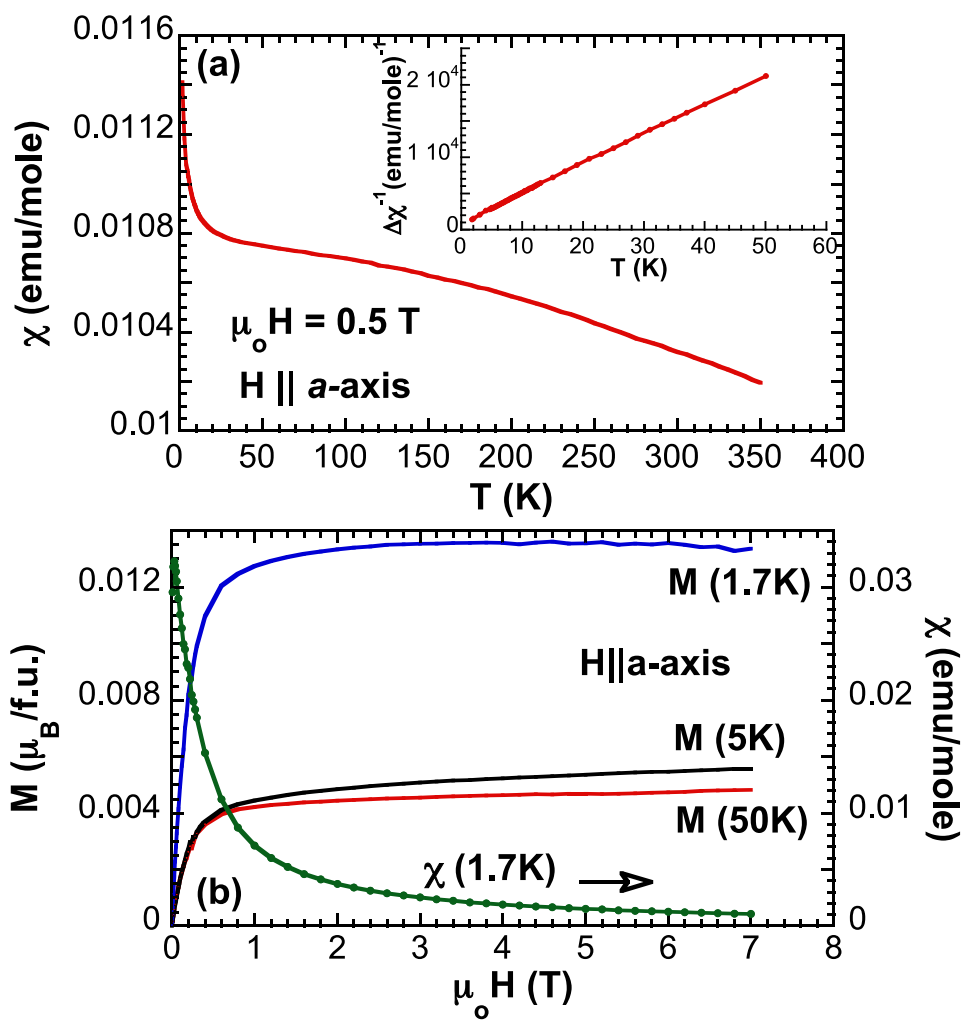

Fig.1

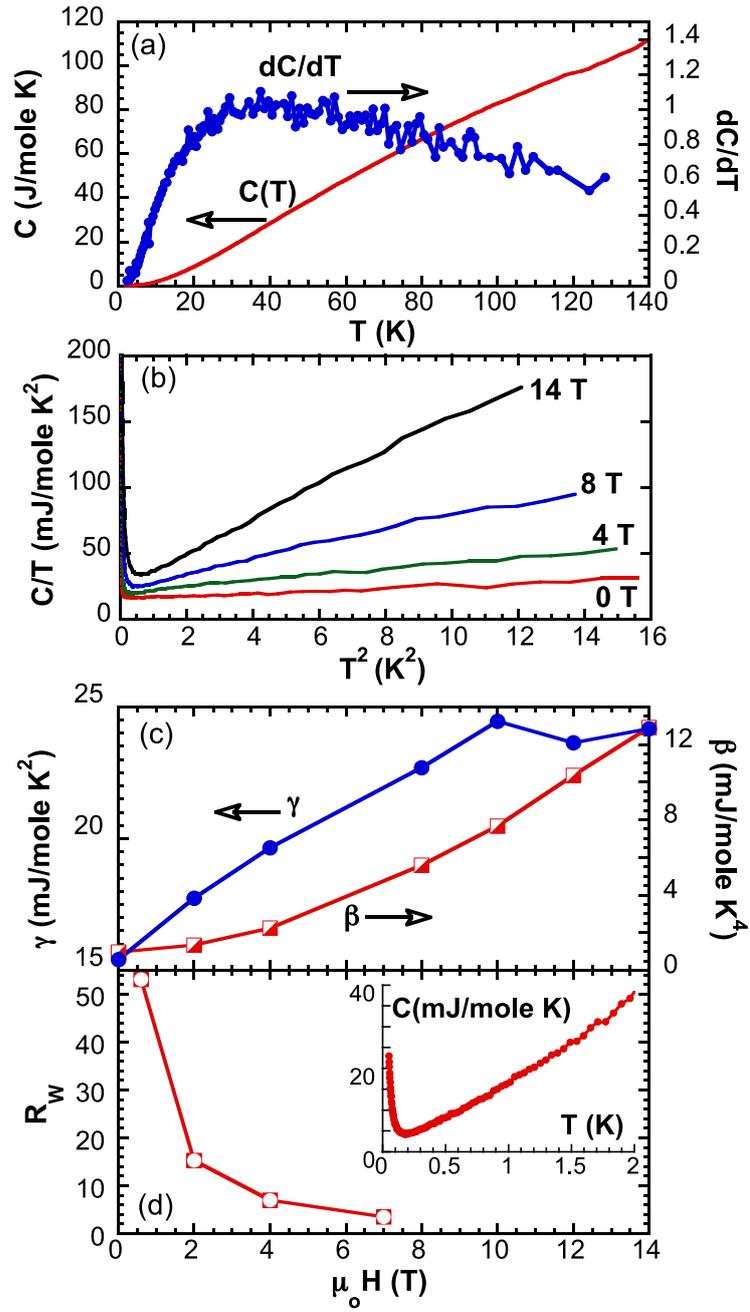

Fig.2

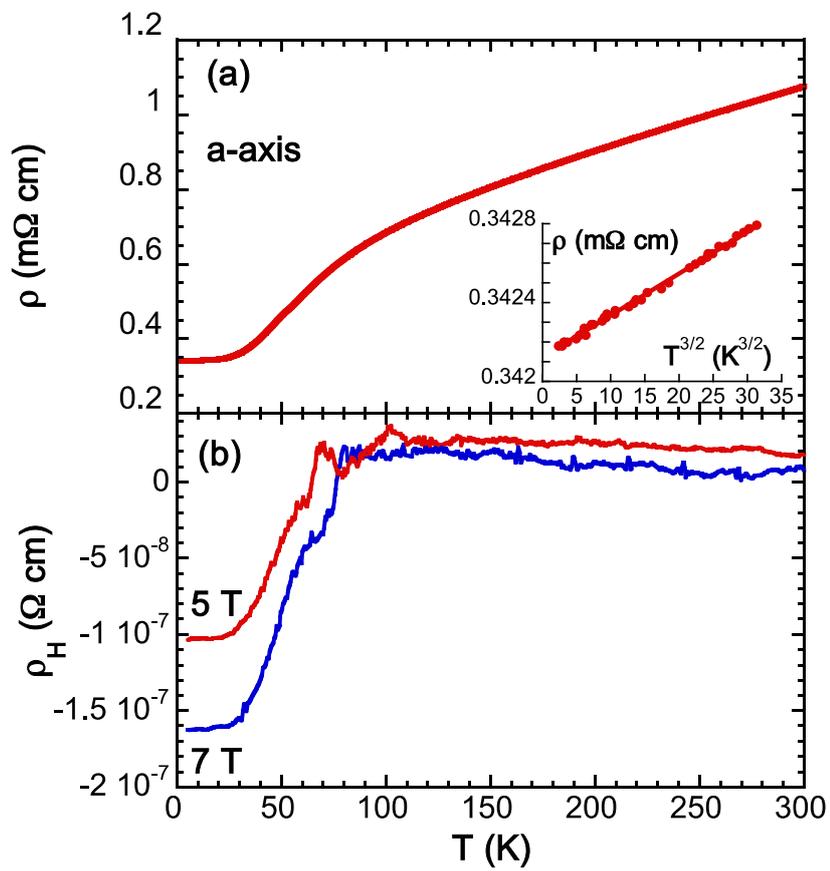

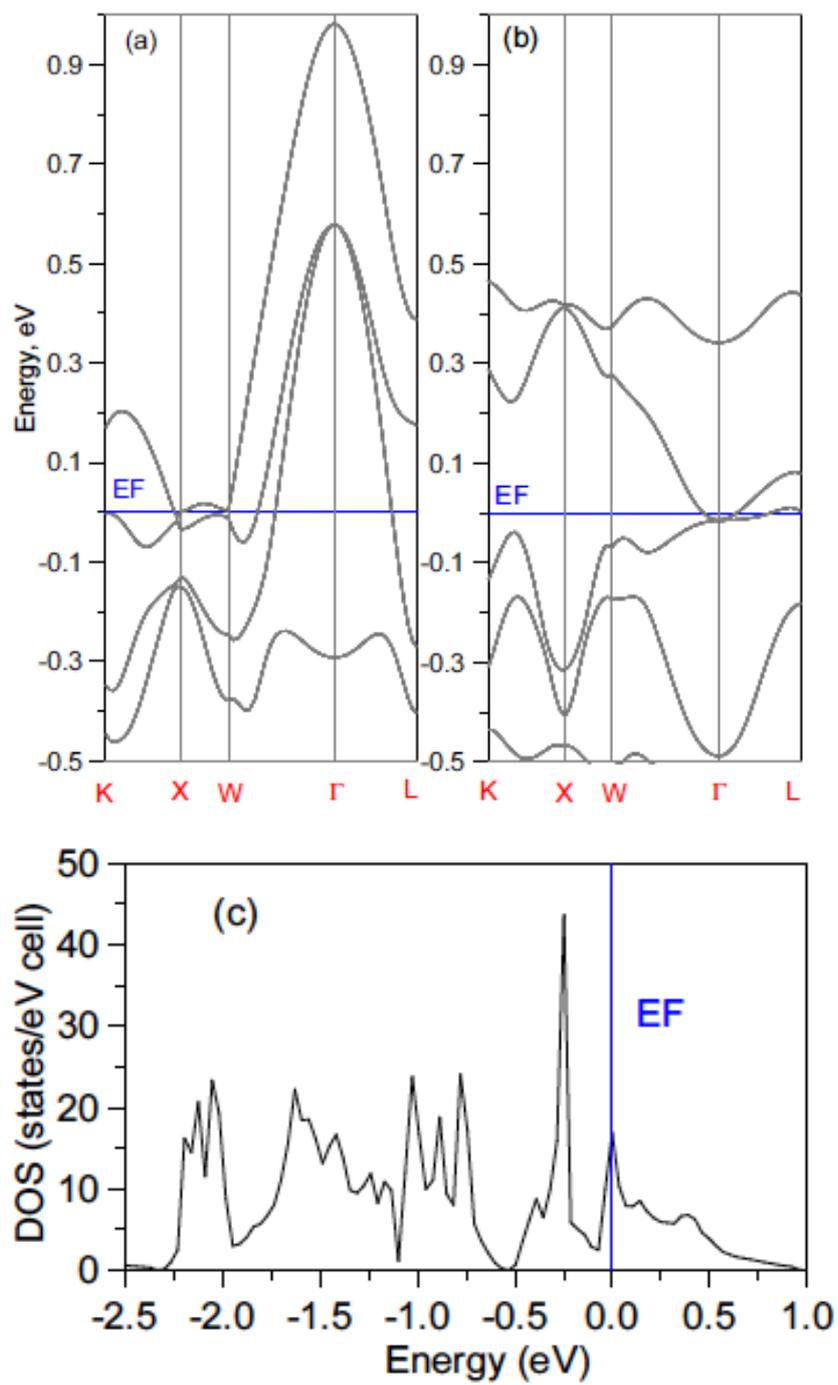

Fig. 4